\documentclass[12pt]{article}
\usepackage{amssymb,slashed,latexsym,ifpdf,amsmath,amsthm}
\usepackage{nicefrac}
\newcommand{\ft}[2]{{\textstyle\frac{#1}{#2}}}

\newsavebox{\uuunit}
\sbox{\uuunit}
    {\setlength{\unitlength}{0.825em}
     \begin{picture}(0.6,0.7)
        \thinlines
        \put(0,0){\line(1,0){0.5}}
        \put(0.15,0){\line(0,1){0.7}}
        \put(0.35,0){\line(0,1){0.8}}
       \multiput(0.3,0.8)(-0.04,-0.02){12}{\rule{0.5pt}{0.5pt}}
     \end {picture}}

\newcommand{\hc}{{\rm h.c.}}
\csname @addtoreset\endcsname{equation}{section}

\newif\ifpdf
\ifx\pdfoutput\undefined
   \pdffalse
   \usepackage{cite}
 \else
   \pdfoutput=1
   \pdftrue
  \usepackage[pdftex]{hyperref}
\fi
   \usepackage[nosort]{cite}
\usepackage{color}
    \definecolor{antiquefuchsia}{rgb}{0.57, 0.36, 0.51}

\newtheorem*{example}{Example}

\def\aD3{{\overline {\rm D3}}}
\def\be{\begin{equation}}
\def\ee{\end{equation}}
\def\ba{\begin{array}}
\def\ea{\end{array}}
\def\bea{\begin{eqnarray}}
\def\eea{\end{eqnarray}}

\def\a{{\alpha}}

\def\b{{\beta}}

\def\be{\begin{equation}}
\def\ee{\end{equation}}
\def\bea{\begin{eqnarray}}
\def\eea{\end{eqnarray}}
\def\ba{\begin{array}}
\def\ea{\end{array}}
\def\bd{\begin{displaymath}}
\def\ed{\end{displaymath}}

\def\a{\alpha}
\def\b{\beta}

\def\d{\delta}
\def\e{\epsilon}           
  
\def\g{\gamma}

\def\k{\kappa}                    

\def\m{\mu}
\def\n{\nu}
  
\def\s{\sigma}                                   

\def\G{\Gamma}

\def\Kahler{K\"{a}hler~}

\def\pa{\partial}                              
\def\>{\rangle} 
\def\<{\langle} 
\def\Dsl{D \hskip-.6em \raise1pt\hbox{$ / $ } }
\def\to{\rightarrow}
\def\pa{\partial}

\newcommand{\eps}{\epsilon}

\def\cn{{\cal N}}

\newcommand{\reef}[1]{(\ref{#1})}

\def\susy{supersymmetry}
\def\susic{supersymmetric}

\begin{document}

\begin{titlepage}
\vspace{.5cm}

\begin{flushright}
 MIT Preprint Number:    \\   MIT-CTP/4829 \\ arXiv:1609.07362
\end{flushright}

\begin{center}
\baselineskip=16pt

\hskip 1cm

\vskip 0.8cm

{\Large {\bf A Geometric Formulation of Supersymmetry}}

\

\

\

{\large  \bf Daniel Z.  Freedman$^{1,2}$}, { \large  \bf Diederik Roest$^3$}, \\

\vskip 0.5 cm

and   {\large  \bf Antoine Van Proeyen$^4$} \vskip 0.8cm
{\small\sl\noindent
$^1$ SITP and Department of Physics, Stanford University, Stanford, California
94305 USA \\\smallskip
$^2$ Center for Theoretical Physics and Department of Mathematics,\\ Massachusetts Institute of Technology, Cambridge, Massachusetts 02139, USA \\\smallskip
$^3$ Van Swinderen Institute for Particle Physics and Gravity,\\ University of Groningen, Nijenborgh 4, 9747 AG Groningen, The Netherlands\\\smallskip
$^4$ KU Leuven, Institute for Theoretical Physics,\\ Celestijnenlaan 200D, B-3001 Leuven,
Belgium}

\end{center}


\vskip 2cm
\begin{center}
{\bf Abstract}
\end{center}

{\small The scalar fields of supersymmetric models are coordinates of a geometric space. We propose a formulation of supersymmetry that is covariant with respect to reparametrizations of this target space.
Employing chiral multiplets as an example, we introduce modified supersymmetry variations and redefined auxiliary fields that transform covariantly under reparametrizations. The resulting action and transformation laws are manifestly covariant and highlight the geometric structure of the supersymmetric theory. The covariant  methods are developed first  for general theories (not necessarily \susic) whose scalar fields are coordinates of a Riemannian target space.}

   \vspace{2mm} \vfill \hrule width 3.cm
{\footnotesize \noindent e-mails:  dzf@math.mit.edu, d.roest@rug.nl, Antoine.VanProeyen@fys.kuleuven.be
 }
\end{titlepage}
\addtocounter{page}{1}
 \tableofcontents{}
\section{Introduction}

This  is the first of two related papers that describe a new covariant formulation of supersymmetry and supergravity. The  multiplets of many \susic{}  theories include scalar fields that span a scalar manifold. Supersymmetry imposes geometric constraints on these manifolds; for example, in ${\cal N} = 1, ~D=4$ supersymmetry, the scalars $z^\alpha$ of chiral multiplets are local coordinates of a K\"{a}hler target space \cite{Zumino:1979et}. The new formulation brings out the geometric nature of such theories.
This means that all quantities we deal with and their symmetry variations  are covariant under reparametrizations of the target space. For the K\"{a}hler manifolds we consider holomorphic  coordinate transformations  $z'^\alpha(z)$. This can be extended to covariance under other target space symmetries in other applications.
Our formulation makes use of both covariant derivatives, and of `covariant symmetry transformations' both for  supersymmetry and other symmetries such as isometries. Moreover,  novel curvature terms appear in several places.

While this geometric framework is quite general, we illustrate it explicitly for the case of ${\cal N}=1$ supersymmetry with chiral multiplets $z^\alpha$, $\chi^\alpha $, $F^\alpha$, consisting of  complex scalar fields, chiral spinors and auxiliary fields. We do not solve the $F^\a$ field equations until interactions are further specified.  This allows the inclusion of constrained chiral multiplets  (some early references are \cite{Volkov:1973ix,Rocek:1978nb,Ivanov:1978mx,Lindstrom:1979kq,Samuel:1982uh,Casalbuoni:1988xh,Komargodski:2009rz,Bergshoeff:2015tra,Hasegawa:2015bza}), which are very useful in applications  of supergravity to inflationary cosmology \cite{Antoniadis:2014oya,Ferrara:2014kva,Kallosh:2014via,Dall'Agata:2014oka,Ferrara:2015tyn}.

In this first paper we illustrate our covariant methods on models with global symmetry.
Our methods are component analogues of the superspace formulations  \cite{Koller:1985kb,Binetruy:2000zx}, which include covariance under coordinate transformations of the target space.
Though many ingredients of our work apply to  ${\cal N}=2$ theories as well,\footnote{The scalars of ${\cal N} = 2$, $D=4$ hypermultiplets parametrize a hyper-K\"{a}hler manifold, and the scalars of vector multiplets determine a rigid special K\"{a}hler manifold. See \cite{Curtright:1979yz,Alvarez-Gaume:1981hm, Sierra:1983cc}.}
we treat only  ${\cal N}=1$ supersymmetry.
 Thus we give an off-shell geometric treatment of the $\cn=1$ nonlinear $\sigma$-model with general K\"{a}hler metric and superpotential.  Conventional interactions with gauge multiplets can be included, but we omit them because special issues associated with target space geometry do not arise.
In addition to super-Poincar\'{e}, one can also recast superconformal theories in the proposed geometric formulation. We will demonstrate in a forthcoming publication \cite{FKRV2} that this allows us to obtain the action and supersymmetry transformations of the fields in the Poincar\'{e} supergravity theory in a straightforward way from those in the parent superconformal theory.

In Section \ref{ss:geomtransf} we develop the geometric formalism at a more  general level, independent of supersymmetry or any specific action. We define covariant concepts and obtain results for covariant symmetry transformations. The geometric ideas  are extended to chiral multiplets in Section \ref{ss:globalsusy}, which includes covariant supersymmetry transformations of the physical and auxiliary fields and the manifestly invariant action.

We hope that the presentation of our ideas in the simpler context of global supersymmetry makes them more accessible to readers and motivates them to read \cite{FKRV2}
where, as we announce in Sec. \ref{ss:conclusion}, the application to the superconformal formalism and then to the physical fields of $\cn =1$ Poincar\'{e} supergravity is given.

\section{Geometrization of transformations}
\label{ss:geomtransf}

\subsection{Scalar fields as coordinate maps}
\label{ss:scalarsonly}

In this subsection we discuss fields that are functions on a Riemannian manifold $M$,   pulled back to spacetime by the coordinate map $\phi^i(x^\mu)$ from spacetime into $M$.   These include scalar and vector fields, $S(\phi)$ and $V^i(\phi)$.  We define the action   of coordinate transformations $\phi^i \rightarrow \phi'^i(\phi)$ on these fields and their derivatives.  We also consider the effects of spacetime symmetries and Killing symmetries that generate isometries of $M$. We denote infinitesimal symmetries generically by $\delta \phi^i$.  The transformations they induce on vectors, called  $\delta V^i$, do not transform as vectors under coordinate transformations. We then propose modified variations, $\hat\delta V^i$, that do transform covariantly. We also define covariant spacetime derivatives $\nabla_\mu V^i$ and discuss their properties.  This material may be familiar to readers, but we present it to set the stage for its extension to the fields of chiral multiplets in Section 3 and the superconformal formulation of supergravity in  \cite{FKRV2}.

Scalar fields correspond to coordinate charts of a manifold and we must allow coordinate reparametrizations, i.e.
\begin{equation}
  \phi ^i \rightarrow  \phi '^i(\phi )\,.
 \label{phirepar}
\end{equation}
We would like to formulate the theory in terms of quantities that take the same form for all choices of coordinates. This requires all quantities to transform covariantly under  (\ref{phirepar}). Thus any scalar quantity $S (\phi)$ should transform as a scalar, while $V^i(\phi)$ should transform as a vector, viz.
 \begin{align} \label{SVdiff}
  S(\phi) &\rightarrow S' (\phi'(\phi ))= S(\phi) \,, \qquad
  V^i(\phi) \rightarrow V'^i(\phi'(\phi )) = \frac{\partial \phi^{\prime i}}{\partial \phi ^j} V^j (\phi) \,.
 \end{align}
Two examples are the scalar $g_{ij} \partial_\mu  \phi^i \partial_\nu \phi^j$ and the vector $\partial_\mu  \phi^i$, where  $g_{ij}(\phi)$ is the metric on $M$, and $\pa_\m$ indicates a spacetime derivative.

We now consider symmetries. Our main interest will be supersymmetry, but the same considerations apply to spacetime and Killing symmetries.
Coordinate covariance must be maintained under any symmetry that the theory might have, and this
requires a careful definition of the action of that symmetry.  The symmetry transformations $\delta\phi^i$ and $\delta\phi'^i$ of coordinate scalars related by the diffeomorphism (\ref{phirepar})  must satisfy
\begin{align}
  \delta \phi'^i &= \frac{\partial \phi^{\prime i}}{\partial \phi ^j} \delta \phi^j \,,
\label{symtrf}
\end{align}
By variation of the first equation in \reef{SVdiff},  one finds that the effect of the symmetry on any scalar quantity is
\begin{align}
  \delta S'(\phi') &= \pa_iS'(\phi')\,\d\phi'^i =\pa_iS'(\phi')\frac{\pa\phi'^i}{\pa\phi^j}\d\phi^j = \pa_j S(\phi)\, \d\phi^j =\delta S (\phi )\,.
\end{align}
This shows that the symmetry variation of any scalar transforms as a scalar under diffeomorphisms.

Varying  the second equation of \reef{SVdiff}, we find (and write in slightly less detail)
\begin{align}
  \delta V'^i & =  \frac{\partial \phi^{\prime i}}{\partial \phi ^j} \delta V^j +\frac{\partial ^2\phi^{\prime i}}{\partial \phi ^j\partial \phi ^k}\, V^j \,\delta \phi ^k\,.
 \end{align}
The second term spoils the desired vectorial property.  However, its structure resembles the transformation law of a connection. This suggests that a \emph{covariant transformation} can be defined using the Christoffel  connection as follows (see \cite{Bergshoeff:2002qk}, \cite[App.14B]{Freedman:2012zz})
\begin{equation}
  \boxed{\hat{\delta }V ^i\equiv  \delta V ^i + \Gamma ^i_{jk}V^j\,\delta \phi ^k\,,}
 \label{hatdeltadefined0}
\end{equation}
for any symmetry operation. Indeed, this definition satisfies the vectorial transformation property
\begin{equation}
   \hat{\delta }V '^i = \frac{\partial \phi ^{\prime i}}{\partial \phi ^j}\hat{\delta }V^j\,.
 \label{covtransfhatdelta0}
\end{equation}

Applying  (\ref{hatdeltadefined0}) to the spacetime translation symmetry $\d\phi^j=\xi ^\mu \pa_\m\phi^j$, we find the
{\it covariant derivative} of a vector quantity $V^i$,  namely
\begin{equation}
  \boxed{\nabla _\mu V^i \equiv \partial _\mu V^i +\Gamma ^i_{jk}V^j\partial _\mu \phi ^k\,.}
 \label{nabladef0}
\end{equation}
The covariant derivative is thus a special case of a covariant symmetry transformation.

For functions that {\it only depend on the coordinate scalars} (and not their spacetime derivatives, or on other fields), there is a simple relation between these covariant operations, viz.
\begin{align}
  \boxed{\nabla_\mu =(\partial _\mu \phi ^j) \nabla _j \,, \qquad \hat{\delta } = (\delta  \phi ^j) \nabla _j \qquad \mbox{on functions of }\phi ^i\,.}
 \label{nablaonscalarfncts}
\end{align}
Note that $\nabla_j$ is the familiar covariant derivative on the target space $M$.

The considerations above can  be generalized to scalars, covectors and tensors in a natural fashion. For scalars, there is no connection term needed and hence no difference between $\delta$ and $\hat \delta$. If an action is constructed as a scalar from vectors and tensors, then invariance under a symmetry operation  is equivalent to invariance under the covariant transformations:
 \begin{align}\label{simp1}
  \delta {\cal L} = {\hat \delta} {\cal L} =0 \,.
 \end{align}
As an example of a tensor, consider the metric $g_{ij}(\phi)$ of which $\Gamma ^i_{jk}$ are the Christoffel symbols. It follows from \eqref{nablaonscalarfncts} that  its covariant transformation under any symmetry vanishes,
\begin{equation}
  \hat{\delta }g_{ij}= (\delta  \phi ^k) \nabla _k g_{ij}=0\,.
 \label{hatdeltag0}
\end{equation}
Simple observations, such as (\ref{simp1}-\ref{hatdeltag0}) will have great practical value in \cite{FKRV2}.

Although the coordinate scalars have a vector index,  they  \emph{do not} transform as vectors as \reef{phirepar} shows.  However, spacetime derivatives of coordinate scalars are vectors. We define
\begin{equation}
  \hat{\delta }\partial _\mu \phi ^i \equiv  \delta \partial _\mu \phi ^i +\Gamma ^i_{jk}\partial _\mu \phi ^j\delta \phi ^k = \nabla _\mu \delta \phi ^i\,.
 \label{hatdeltadmuphi}
\end{equation}
The final equality follows because ordinary derivatives and transformations commute by definition: $\delta \partial_\mu = \partial_\mu \delta$, and it defines a covariant derivative because $\d\phi^i$ transforms as a vector as shown by \reef{symtrf}.

If $\delta \phi ^i$ is only a function of the scalars, we may use the first of (\ref{nablaonscalarfncts}) and write
\begin{equation}
  \hat{\delta } (\partial _\mu \phi ^i) = \nabla _\mu (\delta \phi ^i) =  (\partial_\mu \phi^j) \nabla_j (\delta \phi^i) \,.
 \label{hatdeltadmuphi2}
\end{equation}
Note, however, that the second equation of (\ref{nablaonscalarfncts}) cannot be applied to $\partial _\mu \phi ^i$, since it is not just a function of scalars. We will consider more general vectors that are not just functions of scalars in Sec. \ref{ss:inclother}.

\begin{example}
As a simple example, consider the kinetic Lagrangian of scalars $\phi ^i$:
\begin{eqnarray}
 {\cal L}_\phi=-\ft12 g_{ij}\partial _\mu \phi ^i\partial ^\mu \phi ^j\,,
 \label{Lexample}
\end{eqnarray}
This is form invariant under any diffeomorphism, which means that
\begin{equation}
  {\cal L}_\phi  = -\ft12 g'_{ij}\partial _\mu \phi'^i\partial ^\mu \phi'^j\,,
\end{equation}
in which we have used the standard tensor transformation property of $g_{ij}$
and  the fact that spacetime derivatives of coordinate scalars transform as vectors.

 Diffeomorphisms are usually not symmetries, but those generated by
Killing vectors are exceptions to this rule.   We consider the infinitesimal diffeomorphism
 \begin{equation}
   \delta \phi ^i = k^i(\phi )\,,\quad \mbox{such that} \quad  \nabla_ik_j +  \nabla_jk_i=0\,.
  \label{Killingsymm0}
\end{equation}
First note that ${\cal L}$ is constructed from vectors and tensors, so that \reef{simp1} is valid. Using covariant transformations, we compute
\begin{align}
  \hat{\delta }{\cal L}_\phi & = - g_{ij}\partial ^\mu \phi ^i \nabla _\mu k^j= - g_{ij}\partial ^\mu \phi ^i\partial _\mu \phi ^k \nabla _k k^j  = - \partial ^\mu \phi ^i\partial _\mu \phi ^k \nabla _k k_i=0\,,
 \label{hatdelLphi}
\end{align}
where we have used (\ref{hatdeltag0}) and (\ref{hatdeltadmuphi2}).
This proves invariance under the covariant transformation provided that the Killing equation is satisfied.  Readers should note that covariant methods provide significant simplifications compared with conventional methods.

Finally we note that each Killing symmetry of the target space metric leads to a conservation law, since the
current $J^{\m} = g_{ij}(\pa^\m \phi^i) k^j $ is conserved when the Euler-Lagrange equations for ${\cal L}_\phi$ are satisfied.
\end{example}

\subsection{Inclusion of other fields}
\label{ss:inclother}

Many theories contain fields other than the coordinate scalars $\phi^i(x)$.  Because our main interest is
\susy~ we consider fermions, beginning with $\chi^i(x)$, namely spinors on spacetime that transform as sections of the tangent bundle on a Riemannian target space $M$.  This means that under the reparametrizations $\phi^i\to \phi'^i(\phi)$:
\begin{equation}
  \chi ^i \rightarrow \chi^{\prime i}=\frac{\partial \phi^{\prime i}}{\partial \phi ^j}\chi ^j\,.
 \label{psireparameters}
\end{equation}
Here, as further in this paper and in most treatments of supersymmetry, these fields $\chi ^i$ and $\phi ^i$ are considered as independent such that $\{\phi ^i,\chi ^i\}$ form a basis of the field space.

Symmetries that leave an action invariant take the form:
\begin{equation}
  \delta \phi ^i(\phi ,\chi )\,,\qquad \delta \chi ^i(\phi ,\chi )\,.
 \label{symmetriesphipsi}
\end{equation}
Consider how this changes under reparametrization:
\begin{align}
  \delta \phi ^{\prime i} & = \frac{\partial \phi ^{\prime i}}{\partial \phi ^j}\,\delta \phi  ^j\,, \qquad
 \delta \chi^{\prime i}  = \frac{\partial \phi^{\prime i}}{\partial \phi ^j}\,\delta \chi ^j+\frac{\partial ^2\phi^{\prime i}}{\partial \phi ^j\partial \phi ^k}\,\chi ^j \,\delta \phi ^k\,.
 \label{deltaprime}
\end{align}
We see that $(\delta \phi )^i$ behaves as a vector, but $(\delta \chi )^i$ does not. As in the previous subsection, we define
\begin{equation}
  \boxed{\hat{\delta }\chi ^i\equiv  \delta \chi ^i + \Gamma ^i_{jk}\chi ^j\,\delta \phi ^k\,,}
 \label{hatdeltadefined}
\end{equation}
for any symmetry operation. This transforms covariantly as the vectorial quantity (\ref{covtransfhatdelta0}). Covariant derivatives (\ref{nabladef0}) are defined in the same way.
These definitions and properties apply also to other vector functions such as $\partial _\mu\phi ^i$.

Since the vectors $\chi^i$ are not functions of the coordinate scalar fields, these covariant expressions for transformations and spacetime derivatives cannot be related to the covariant target space derivative: the relations \eqref{nablaonscalarfncts} are valid for quantities that depend only on the coordinate scalars.

The conventional Ricci identity for covariant target-space derivatives:
 \begin{align}
  [ \nabla_i , \nabla_j ] V^k = R_{ij}{}^k{}_l V^l \,. \label{curvature}
 \end{align}
is a fundamental relation in Riemannian geometry and may be viewed as the definition of the curvature tensor.
We now derive analogous relations for the commutators of covariant spacetime derivatives and transformations.

We begin with $[\nabla_\m,\nabla_\n]V^k(x)$ for any vector function.  In full detail the quantity $\nabla_\mu \nabla_\nu V^k$ contains many terms, but it can be simplified by omitting terms that serve
only to covariantize the final result and hence are inessential. The essential terms are
\begin{align}
  \nabla_\mu \nabla_\nu V^k = \partial_\mu \partial_\nu V^k + (\partial_\mu \Gamma^k_{ij}) (\partial_\nu \phi ^i) V^j + \ldots\,,
\label{essential}
 \end{align}
where we omitted terms with explicit (non-differentiated) Christoffel symbols. Since the Christoffel symbol itself only depends on scalar fields,  we can apply the chain rule to its space-time derivative. Therefore we find
\begin{align}
 [ \nabla_\mu , \nabla_\nu ] V^k = 2 \partial_{[\ell} \Gamma^{k}_{i]j}  (\partial_\mu \phi^\ell)(\partial_\nu \phi^i) V^j + \ldots \,,
 \end{align}
whose covariant form is
\begin{equation}
 [ \nabla_\mu , \nabla_\nu ] V^k  = R_{ij}{}^k{}_\ell (\partial_\mu \phi^i) (\partial_\nu \phi^j) V^\ell \,,
 \label{commcovder}
 \end{equation}
valid for any vector $V^k$.

There is a similar relation for  the commutator of two covariant symmetry transformations. Suppose we have symmetries $\d(\eps)V^i$ and their covariant forms $\hat\delta (\eps) V^i$ from \eqref{hatdeltadefined} with  symmetry parameters $\eps$.
Their commutators covariantize as (with $\delta _1=\delta (\epsilon _1)$)
\begin{equation}
  \left[\delta_1,\delta_2\right] V^i=\delta_3 V^i\  \rightarrow \ \left[\hat{\delta}_1,\hat{\delta}_2\right] V^i=\hat{\delta }_3 V^i + R_{k\ell}{}^i{}_j V^j(\delta_1\phi ^k)( \delta  _2\phi ^\ell)\,,
 \label{commcovdelta}
\end{equation}
where $\delta _3$ is determined in terms of $\epsilon _1$ and $\epsilon _2$ by the structure functions of the symmetry.\footnote{For fermions of the $\cn=1$ K\"{a}hler $\s$-model, an analogue of   \eqref{commcovdelta} was derived in \cite[(14.118)]{Freedman:2012zz}.}
To prove this we use the shortcut of  \eqref{essential} and write
\begin{align}
\hat\delta_1\hat\delta_2 V^i =\delta_1\delta_2 V^i +\d_1\G^i_{jk}V^j\delta_2\phi^k+\ldots\,.
\end{align}
We proceed as above to obtain \eqref{commcovdelta}.

There is an analogous relation for $(\hat\delta\nabla_\mu-\nabla_\mu\hat\delta)V^i.$  To prove it, realizing that a covariant expression must arise at the end, we calculate
\begin{align}
  \hat{\delta }\nabla _\mu V^i & = \delta \nabla _\mu V^i+\ldots = \delta \partial _\mu V^i + \delta \Gamma ^i_{jk}V^j \partial _\mu \phi ^k+\ldots = \delta \partial _\mu V^i + \partial _\ell\Gamma ^i_{jk} \, \delta \phi ^\ell\,V^j \partial _\mu \phi ^k+\ldots\,,\nonumber\\
  \nabla _\mu \hat{\delta }V^i & = \partial _\mu \hat{\delta }V^i+\ldots = \partial _\mu \left(\delta V^i +\Gamma ^i_{jk}V^j\delta \phi ^k\right)+\ldots = \partial _\mu \delta V^i + \partial _\ell\Gamma ^i_{jk}V^j\partial _\mu \phi ^\ell \delta \phi ^k+\ldots\,,
 \label{proofcomm}
\end{align}
After combining these two expressions, we find the covariant relation
\begin{align}
\hat{\delta }\nabla _\mu V^i = \nabla _\mu \hat{\delta }V^i+ R_{k\ell}{}^i{}_jV^j(\delta \phi ^k)(\partial _\mu \phi ^\ell) \,. \label{commcovderdelta}
\end{align}

\begin{example}
As an extension of the previous example, consider the Lagrangian of Majorana fermion fields $\chi ^i$, with Lagrangian
\begin{eqnarray}
 {\cal L}_\chi= -\ft12 g_{ij}\bar \chi ^i \gamma ^\mu \nabla _\mu \chi ^j\,,\qquad \quad\nabla_\m \chi^i\equiv \pa_\mu \chi^i+\G^i_{jk}\pa_\m\phi^j\chi^k\,.
 \label{Lchiexample}
\end{eqnarray}
We consider a Killing symmetry acting on both bosons and fermions, for which
\begin{equation}
  \delta \phi ^i = k^i(\phi )\,,\qquad \delta \chi ^i = \chi ^j\partial _j k^i\,.
 \label{Killingsymm}
\end{equation}
Again we will use covariant methods to demonstrate the invariance of the action.
First we use \eqref{hatdeltadefined} to define a covariant Killing transformation of the fermions:
\begin{equation}
  \hat{\delta }\chi ^i = \chi ^j\nabla  _j k^i\,.
 \label{hatdeltachi}
\end{equation}
Then, (\ref{commcovderdelta}) implies
\begin{eqnarray}
 \hat{\delta }\nabla  _\mu \chi ^i &=& \nabla _\mu \left(\chi ^j\nabla  _j k^i\right)+ R_{k\ell}{}^i{}_j\chi ^j k ^k(\partial _\mu \phi ^\ell)\nonumber\\
 &=&\nabla  _j k^i \nabla _\mu \chi ^j +\left(\nabla _\ell \nabla  _j k^i+R_{k\ell}{}^i{}_j k ^k\right)\, \chi ^j\partial _\mu \phi ^\ell   \,.
 \label{Killinghatdelnabla}
\end{eqnarray}
Using (\ref{Killingsymm0}), the last part vanishes:
\begin{align}
  \nabla _\ell & \nabla  _j k^i+R_{k\ell}{}^i{}_j k ^k =\ft12\left[\nabla _\ell,\, \nabla  _j\right] k^i + \nabla _{(\ell} \nabla  _{j)} k^i+R_{k\ell}{}^i{}_j k ^k\nonumber\\
  &=\ft12R_{\ell j}{}^i{}_k k^k-\nabla _{(\ell}\nabla^i k_{j)}+R_{k\ell}{}^i{}_j k ^k =\left[\ft12R_{\ell j}{}^i{}_k-R_{(\ell}{}^i{}_{j)k}+R_{k\ell}{}^i{}_j\right]k^k =0\,,
 \label{vanishingRs}
\end{align}
using the Bianchi identity. Hence, we have
\begin{equation}
\hat{\delta }\nabla  _\mu \chi ^i  = \nabla _\mu \chi ^j \nabla  _j k^i\,,
 \label{Killinghatdchi}
\end{equation}
after which the invariance follows in the same way as in (\ref{hatdelLphi}).
\end{example}

\section{Global supersymmetry}
\label{ss:globalsusy}

In this section we present the covariant formulation of ${\cal N} =1$ globally supersymmetric theories, restricting to chiral multiplets. The scalar geometry in this case is a \Kahler manifold, spanned by (anti-)holomorphic coordinates $\{  Z^\alpha,~\bar Z^{\bar \beta} \} $. Its metric is determined by the \Kahler potential $K(Z,\bar Z)$ and given by
 \begin{equation}
  g_{\alpha \bar \beta }=K_{\alpha \bar \beta }\equiv\partial _\alpha \partial _{\bar \beta }K(Z,\bar Z)\,.
 \label{defgalphabeta}
 \end{equation}
The other fields in the multiplets are fermions\footnote{We use the notation that $\chi ^\alpha$ and $\chi ^{\bar \alpha}$ are left- and right-handed, respectively, i.e.  $\chi^\a=P_L\chi^\a,~
\chi ^{\bar \alpha}=P_R\chi ^{\bar \alpha}$.} $\chi^\alpha$, auxiliary fields $F^\alpha$ and their conjugates $\chi^{\bar\a}$, $\bar F^{\bar\a}$. The early stages of our discussion are similar to Appendix 14B of \cite{Freedman:2012zz}, but  we develop the covariant treatment more completely here.

\subsection{Supersymmetry transformations}
\label{ss:susytrchiral}

The conventional SUSY transformation rules of a chiral multiplet are
 \begin{align} \label{delta}
  \boxed{\delta Z^\alpha = \frac{1}{\sqrt{2}} \bar \epsilon \chi^\alpha \,,}\qquad
  \delta \chi^\alpha = \frac{1}{\sqrt{2}} P_L \left(\slashed{\partial } Z^\alpha + F^\alpha \right) \epsilon \,,\qquad
  \delta F^\alpha = \frac{1}{\sqrt{2}} \bar \epsilon \slashed {\partial} \chi ^\alpha  \,.
\end{align}
Only the $\d Z^\a$ transformation is covariant under reparametrizations of the coordinate fields
\begin{equation}\label{reparams}
Z^\a\rightarrow Z'^\a(Z^\b)\,.
\end{equation}
This follows because both sides of the equation transform as tangent vectors, i.e.
\begin{equation}
\d Z^\a\rightarrow \delta Z'^\a= \frac{\pa Z'^\a}{\pa Z^\b} \d Z^\b\qquad\quad \chi^\a \to \chi'^\a =  \frac{\pa Z'^\a}{\pa Z^\b}\chi^\b\,.
\end{equation}

By contrast, $\delta \chi^\alpha$ does not transform as a vector.
This can be checked explicitly, as in (\ref{deltaprime}):
\begin{equation}
  \delta \chi '^\alpha =\delta \left( \frac{\pa Z'^\a}{\pa Z^\b} \chi^\beta \right)=\frac{\pa Z'^\a}{\pa Z^\b} \frac{1}{\sqrt2} P_L\left(\slashed{\partial}Z^\beta  + F^\beta \right)\epsilon +\chi^\beta \frac{\partial ^2 Z'^\alpha }{\partial Z^\beta \partial Z^\gamma }
  \frac{1}{\sqrt2} \bar \epsilon \chi ^\gamma \,.
 \end{equation}
 If we want to identify this with the transformed variables
\begin{equation}
  \delta \chi '^\alpha = \frac{1}{\sqrt2} P_L(\slashed{\partial}Z'^\alpha + F'^\a)\epsilon \,,
 \label{transfvarchi}
\end{equation}
we find (after a Fierz rearrangement) that $F^\a$ is not a vector. Rather it undergoes the non-covariant transformation
\begin{equation}
F'^\alpha = \frac{\pa Z'^\a}{\pa Z^b}F^\b -\frac12 \frac{\pa^2 Z'^\a}{\pa Z^\b \pa Z^\g} (\bar \chi^\b \chi^\g)\,.
\label{noncovtransF}
\end{equation}
To repair the situation we proceed in two steps. First, as in (\ref{hatdeltadefined}), we must define  the covariant transformation
\begin{equation}
  \hat{\delta }\chi ^\alpha \equiv \delta \chi ^\alpha + \Gamma ^{\alpha}_{\beta \gamma}\chi ^\beta \delta Z^\gamma \,.
 \label{defhatdeltaOmega0}
\end{equation}
This leads to
\begin{equation}
  \hat{\delta } \chi ^\alpha = \frac{1}{\sqrt2} P_L\left({\slashed{\partial }}Z^\alpha + F^\alpha \right) \e -\frac1{2\sqrt{2}} \Gamma ^{\a}_{\b \g}P_L\epsilon \,(\bar \chi ^\b \chi ^\g )  \,.
 \label{hatdelOmega0}
\end{equation}
From the general properties of $\hat \delta$ it follows that the left hand side should be a coordinate vector. Since $\partial _\mu Z^\alpha $ is a vector, the other two terms together should transform as a vector. We therefore define the covariant counterpart of the auxiliary fields as\footnote{This variable has already been introduced in \cite{Kugo:1982mr} to simplify solutions of the field equations.}
\begin{equation}
  \hat F^\alpha = F^\alpha - \frac12 \G^\alpha_{\beta \gamma}(\bar \chi^\beta \chi^\gamma) \,.
 \label{covariantF0}
\end{equation}
Note that $\hat F^\alpha$ is covariant, although $ F^\alpha$ is not, because the transformation of  the added fermion bilinear cancels the non-covariant part of (\ref{noncovtransF}). Then the covariant form of the transformation, consisting of manifestly covariant quantities, reads \cite{Freedman:2012zz}
\begin{align}
 \boxed{ \hat{\delta} \chi^\alpha =  \frac{1}{\sqrt{2}} P_L \left(\slashed{\partial} Z^\alpha+\hat F^\alpha \right)\epsilon \,. \label{finalcovtrOmega0}}
\end{align}
Note that this requires covariantization of both the SUSY transformations as well as the auxiliary fields.

The covariant version of the auxiliary field  transformation rule must still be found.  We begin by calculating
\begin{align}
\d \hat F^\alpha =\frac{1}{\sqrt{2}}\bar\epsilon\bigg[\slashed{\partial} \chi^\alpha -\frac12 (\chi^{\bar \d} \pa_{\bar \d}\G^\a_{\b \g} +\chi^\d \pa_\d \Gamma ^\a_{\b \g})(\bar\chi^\b \chi^\g)
+\Gamma ^\a_{\b \g}(\slashed{\pa}Z^\b - F^\b)\chi^\g \bigg] \,.
\label{delhatF0}
\end{align}
The $\pa_{\bar \d}\Gamma ^\a_{\b\g}$ is the Riemann tensor of the K\"{a}hler manifold.
We then use
\begin{equation}
  \partial _\d\Gamma^\alpha _{\b \g}= \partial _\d \left(g^{\a \bar \k}K_{\b \g \bar \k}\right)= -\Gamma ^\a_{\d \k}\Gamma ^\k_{\b \g}+ g^{\a \bar \k} K_{\d \b \g \bar \k}\,.
 \label{workondG0}
\end{equation}
The last term is symmetric in $(\d \b \gamma )$ and its contribution to (\ref{delhatF0}) vanishes due to index contraction with the three left-handed chiral fermions.
(Fierz rearrangement is one way to show this.)
The $\Gamma\Gamma$ structure combines with the $\G F$ term to produce  $\Gamma \hat{F}$. Hence we have
\begin{equation}
  \delta \hat{F}^\alpha = \frac{1}{\sqrt{2}} \bar \epsilon \left[\slashed{\partial }\chi ^\alpha +\ft12 R_{\b \bar \d}{}^\a{}_\g \chi ^{\bar \d}\overline{\chi} ^\b \chi^\g +\Gamma ^\a_{\b\g}(\slashed{\partial} Z^\b -\hat{F}^\b)\chi^\g \right]\,.
 \label{resultdelhatF0}
\end{equation}
The $\slashed{\partial}Z^\beta $ remains, and its role is to turn $\slashed{\partial}\chi^\alpha$ into the reparametrization covariant  derivative of $\chi^\a$ (as in  \eqref{Lchiexample})
\begin{equation}
  \nabla_\mu  \chi ^\a= \pa_\mu \chi ^\alpha +\Gamma ^\a_{\b\g}\chi ^\b{\pa}_\mu Z^\g\,.
 \label{wcD0}
\end{equation}
The last term in (\ref{resultdelhatF0}) disappears when we use $\hat\d \hat F^\a= \d\hat F^\a + \G^\a_{\b\g}\d Z^\b \hat F^\g$ to form the manifestly covariant transformation
\begin{equation}\label{delFcov0}
\boxed{\hat{\delta }\hat F^\a=\frac{1}{\sqrt{2}}\bar\epsilon\left[ \slashed {\nabla} \chi ^\alpha +\ft12R_{\b \bar \d}{}^\alpha {}_\g \chi ^{\bar \d}\overline{\chi}{} ^\b\chi^\g\right]\,.}
\end{equation}
Note that, by replacing both the quantities and transformations with their covariant counterparts, we have gained a covariantization and a curvature term with respect to the conventional formulation.

This is consistent with the covariant superalgebra (\ref{commcovdelta}), which applied on $\chi ^\alpha$ implies
\begin{align}
 \left[\hat{\delta}_1,\hat{\delta}_2\right]\chi ^\alpha &=\hat{\delta }_3 \chi ^\alpha + R_{\g\bar \d}{}^\a{}_\b\chi ^\b\left[\ft12\bar \epsilon _1\chi ^\d\,\bar \epsilon _2\chi ^{\bar \g}- (1\leftrightarrow 2))\right]\nonumber\\
  & =  \hat{\delta }_3\chi ^\alpha -\ft14 \left[R_{\g\bar \d}{}^\a{}_\beta P_L\epsilon _1(\bar \chi ^\b  \chi ^\g)\bar \epsilon _2\chi ^{\bar \d} - (1\leftrightarrow 2))\right]\,.
 \label{explicitcommOmega0}
\end{align}
The first term is a covariant translation:
\begin{align}
 \hat \delta_3\chi ^\alpha=\xi ^\mu \nabla _\mu \chi ^\alpha\,,
\end{align}
where $\xi ^\mu $ is determined by the parameters of the supersymmetry transformations 1 and~2: $\xi ^\mu = \tfrac12 \bar \epsilon_2 \gamma^\mu \epsilon_1$.
The second term appears in the explicit calculation of commutators of covariant transformations due to applying  (\ref{delFcov0}).

\subsection{Supersymmetric action}

The same covariant methods apply to the globally supersymmetric action for chiral multiplets. Terms of the $\s$-model kinetic Lagrangian (called the $D$-term in \cite[(14.15)]{Freedman:2012zz})
 combine into the simpler structure
\begin{equation}
  \boxed{ [K ]_D = K_{\a \bar \b}\left[ -{\pa} _\mu Z^\alpha {\pa}^\mu \bar Z^{\bar \b}-\frac12\bar \chi^\alpha {\slashed{\nabla }}\chi^{\bar \b}-\frac12\bar \chi^{\bar \b}
\slashed{\nabla}\chi^\alpha +\hat F^\alpha \bar {{\hat F}}^{\bar \b } \right] +\,\frac14R_{\a \bar \g \b \bar \d}\,\bar \chi^\alpha \chi^\b \bar \chi^{\bar \g}\chi^{\bar \d} \,,}
 \label{kinchiralConf0}
\end{equation}
Similarly, the superpotential term or $F$-term is given by
\begin{equation}
 \boxed{  [{W}]_F=
{W}_\a \hat F^\alpha -\ft12 \nabla_\alpha { W}_{\b} \bar \chi^{\a} \chi^{\b} + \hc\,.}
 \label{WF0}
\end{equation}
Note that these formulae contain only covariant quantities. For instance, the term with four derivatives of the K\"{a}hler potential in the conventional formula has combined into the Riemann curvature tensor of the K\"{a}hler manifold. Similarly, the superpotential becomes covariant when the $\hat F^\alpha $ variable is used.

The proof that (\ref{WF0}) is invariant under the covariant SUSY transformations is to a large extent identical to
the proof with ordinary transformations. Extra terms appear in two places, both due to covariantization. First, the last term in the variation of the auxiliary field, (\ref{delFcov0}),
leads to a contribution of the transformation of (\ref{WF0}) of the form
\begin{equation}
  \hat{\delta }\left({W}_\alpha \hat F^\alpha\right)=\frac{1}{2\sqrt{2}}{W}_\alpha R_{\beta \bar \delta }{}^\alpha{}_\gamma  \bar\epsilon\chi ^{\bar \delta }\overline{\chi}{} ^\beta\chi^\gamma+\ldots \,.
 \label{RtransfW0}
\end{equation}
The second contribution comes from the transformation of the covariant derivative $\nabla_\alpha {W}_{\beta}$. In contrast to the ordinary proof, where $\partial _\alpha {W}_\beta $ appears, which is holomorphic, this contains also the antiholomorphic fields and thus, according to (\ref{nablaonscalarfncts}), which is applicable since the argument depends only on the scalars:
\begin{equation}
  \hat{\delta }\nabla _\alpha {W}_\beta = \delta Z^\gamma\nabla _\gamma\nabla _\alpha {W}_\beta  + \delta \bar Z^{\bar \gamma}\nabla _{\bar \gamma}\nabla _\alpha{W}_\beta \,.
 \label{hatdelWIJ0}
\end{equation}
The first term, as in the usual proof, does not contribute due to a Fierz identity and the symmetry $(\alpha \beta \gamma )$. The second term gives rise to a another curvature term since $\nabla _{\bar \gamma}{W}_\beta=0$:
\begin{equation}
  \hat{\delta }\nabla _\alpha {W}_\beta =\delta Z^\gamma\nabla _\gamma\nabla _\alpha{W}_\beta - R_{\bar \gamma \alpha}{}^\delta{}_\beta\,{W}_\delta\,\delta Z^{\bar \gamma}\,.
 \label{hatdelWIJ1}
\end{equation}
This gives thus the contribution to the transformation of (\ref{WF0}) of the form
\begin{equation}
 \hat{\delta }\left( -\ft12{W}_{\alpha ;\beta}\,\bar \chi^\alpha \chi ^{\beta}\right)= \ldots + \frac{1}{2\sqrt{2}}\, R_{\bar \gamma \beta}{}^\delta {}_\alpha{W}_\delta \bar\epsilon\chi ^{\bar \gamma}\,\bar \chi^{\alpha }\chi ^{\beta}+\ldots \,,
 \label{contibdelW20}
\end{equation}
which cancels with (\ref{RtransfW0}), completing the proof.

Similarly, the $D$-term is invariant under the covariant supersymmetry transformations. Again this calculation mainly entails a covariantized version of the usual proof of invariance; only at a very few places do explicit curvature terms arise. This includes the variation of the explicit curvature term that is quartic in fermions, which gives
 \begin{align}
  \hat \delta & (\tfrac14 R_{\alpha \bar \gamma \beta \bar \delta} \bar \chi^\alpha \chi^\beta \bar \chi^{\bar \gamma} \chi^{\bar \delta} )
   = \frac{1}{2\sqrt2} R_{\alpha \bar \gamma \beta \bar \delta} \bar \chi^\alpha (\slashed \partial Z^\beta + \hat F^\beta) \epsilon \bar \chi^{\bar \gamma} \chi^{\bar \delta} + \hc
 \end{align}
The term involving the auxiliary field in this expression is canceled by the SUSY variation of the explicit auxiliary field in the action. Similarly, the remaining term is opposite in sign to the variation of the kinetic terms of the fermions, for which it is convenient to use \eqref{commcovderdelta}.

The variation of the curvature term also contains a quintic term in the fermions proportional to
\be
[\nabla_\zeta R_{\alpha \bar \gamma \beta \bar \delta} +\nabla_\alpha R_{\zeta \bar \gamma \beta \bar \delta}]
(\bar\epsilon \chi^\zeta)(\bar \chi^\alpha\chi^\beta)\ldots\,,
\ee
where we have used the Bianchi identity to symmetrize in $\zeta \leftrightarrow \alpha$. However, a Fierz rearrangement can be used to show that this expression is equal to $(-1/2)$ of itself and therefore vanishes,
thus completing the proof of \susy.

\subsection{On shell transformation of auxiliary fields}

The covariant transformation $\hat\d\hat F^\a$ in \eqref{delFcov0} is an off-shell relation.  It was found by covariantizing the standard rule in \eqref{delta},  and it is independent of the particular \susic \, action that determines the dynamics of the system.  If that action is chosen as the integral of $[K]_D+[W]_F$ for   the $\cn=1$ nonlinear $\s$-model,  then $\hat F^\a$ is an auxiliary field.  It can be eliminated by its equation of motion to obtain the on-shell value 
 \begin{align}
 \hat F^{\alpha} =- K^{\a\bar\b}\pa_{\bar\b}\overline W \equiv - \overline{ W}^{\alpha} \,.
 \end{align}
This is a function of the coordinate scalars, so its SUSY variation is
\begin{align}\label{delFonshell}
\d_{\text{on-shell}} \hat F^{\alpha}= - \frac{1}{\sqrt{2}}  \nabla_{\bar \beta} \overline{W}^\alpha \bar \epsilon \chi^{\bar \beta} \,.
\end{align}
Consistency requires that the off-shell version reduce to this when the equations of motion of other fields of the system are satisfied. Indeed we can rewrite  (\ref{delFcov0}) as follows:
\begin{align}\label{delFcovDiederik}
\hat{\delta }\hat F^\alpha = & \frac{1}{\sqrt{2}}\bar\epsilon\left[ \gamma^\mu\hat{\nabla }_\mu\chi^\alpha +\ft12R_{\beta \bar \delta}{}^\alpha{}_\gamma \chi^{\bar \delta}\overline{\chi} ^\beta \chi^\gamma
+  \nabla_{\bar \beta} \overline{W}^\alpha \chi^{\bar \beta} \right]  - \frac{1}{\sqrt{2}}  \nabla_{\bar \beta} \overline{W}^\alpha \bar \epsilon \chi^{\bar \beta} \,.
\end{align}
The term in brackets $\left[\dots\right]$ vanishes by the fermion equation of motion, so the consistency test is satisfied.

\section{Conclusion}
\label{ss:conclusion}

The symmetry transformations of any theory in which the scalar fields are coordinates of a geometric target space can be expressed in terms of quantities that are covariant under target space reparametrizations. We have discussed the general rules  to construct covariant transformations and derivatives for scalars and other fields, and shown that  their commutators involve novel curvature terms.  It should be emphasized that much of this covariant structure is independent of any dynamical model.  As a particular application, we applied this method to the chiral multiplets in ${\cal N}=1$, $D=4$ supersymmetry and highlighted how all fields, transformation laws and the action can be written in covariant form under K\"{a}hler reparametrizations.

These intrinsically elegant techniques have wide applications. In the second paper of this series (\cite{FKRV2}, in preparation) we will apply them to the dynamics of chiral multiplets in a general $\cn=1$ supergravity theory. In that example  covariant methods are applied first at the superconformal level and it is shown how they simplify the passage to the embedded projective K\"{a}hler manifold of an off-shell version of the physical supergravity theory that is invariant under Poincar\'{e} supersymmetry.

\medskip
\section*{Acknowledgements}

\noindent The authors warmly thank Renata Kallosh for suggesting this topic and for participating in the early stages of our research.

The research of DZF is partly supported by the US National Science Foundation grant NSF PHY-1620045.
The work of AVP is supported in part by the Interuniversity Attraction Poles Programme
initiated by the Belgian Science Policy (P7/37).
This work is supported in part by the COST Action MP1210 `The
String Theory Universe'.

We acknowledge hospitality of the Department of Physics of Stanford University during a visit in which this work was initiated, and of the GGI institute in Firenze, where this work was completed during the workshop `Supergravity: what next?'.


\providecommand{\href}[2]{#2}\begingroup\raggedright\endgroup

\end{document}